\documentclass[journal]{IEEEtran}
\usepackage{xcolor}
\usepackage{stmaryrd}
\usepackage{amsmath}
\usepackage{amsthm}
\usepackage{xfrac}
\usepackage{amssymb}
\usepackage{psfrag}
\usepackage{graphicx}
\usepackage{epstopdf}
\usepackage{todonotes}
\usepackage{cite}
\usepackage{algorithmic}
\usepackage{algorithm}
\usepackage{svg}
\usepackage{mathtools}
\usepackage{hyperref}
\usepackage{hyphenat}
\usepackage{booktabs}
\usepackage{accents}
\usepackage{bbm}
\usepackage{comment}
\usepackage{balance}
\usepackage{mathtools,amssymb,lipsum}

\usepackage{cuted}
\newlength{\dhatheight}

\usepackage{float}
\usepackage{multirow}

\usepackage[caption=false,font=footnotesize]{subfig}

\usepackage{xparse,mathtools}

\begin{document}
%

\title{A Theoretical Framework for Frequency-Shift Filtering-Based Interference Mitigation for OFDM Systems}

\author{Xinrui Li, Karim Said, Daniel J. Jakubisin,~\IEEEmembership{Senior Member,~IEEE}, R. Michael Buehrer,~\IEEEmembership{Fellow,~IEEE}, \\and               Jeffrey H. Reed,~\IEEEmembership{Life Fellow,~IEEE}
\thanks{This work was supported in part by the U.S. Department of Commerce’s National Telecommunications and Information Administration (NTIA) under the Public Wireless Supply Chain Innovation Fund Grant Program (Award 24-60-IF2415: ASPEN - Advanced Signal Processing Enhancement for Next-Generation Open Radio Units), administered by the National Institute of Standards and Technology. }
\thanks{The authors are with the Wireless@VT, Bradley Department of ECE, Virginia Tech, Blacksburg, VA 24061 USA, E-mail: (lxinrui3@vt.edu; karim5@vt.edu; rbuehrer@vt.edu; reedjh@vt.edu).}
\thanks{Daniel J. Jakubisin is also with the Hume Center for National Security and Technology, National Security Institute, Virginia Tech, Blacksburg, VA 24061 USA, Email: (djj@vt.edu). }
}
%




\maketitle
\begin{abstract}
Orthogonal frequency division multiplexing (OFDM) is widely adopted in modern wireless systems due to its high spectral efficiency and robustness to multipath propagation. However, its performance can be severely degraded by narrowband and wideband interference, particularly in spectrally congested environments where interference overlaps significantly in time and frequency with the desired signal. In this paper, we investigate interference mitigation for OFDM signals using the FREquency-SHift (FRESH) filter, which exploits cyclostationarity inherent to the waveform or induced by the modulation design. By leveraging correlations between frequency-shifted versions of the received signal, the FRESH filter provides optimal linear filtering for the cyclostationary processes. We first analyze the second-order statistics of OFDM signals with known structure and characterize their cyclostationary properties. Based on this analysis, we derive a compact closed-form matrix expression for the optimal FRESH filter coefficients under the minimum mean-square error (MMSE) criterion. The proposed framework enables offline computation of the filter coefficients when the signal structure is known and interference parameters are estimated, thereby reducing real-time implementation complexity. Numerical results demonstrate the effectiveness of the proposed method in suppressing interference and improving detection performance compared to conventional approaches.
\end{abstract}


\section{Introduction}
Orthogonal frequency division multiplexing (OFDM), widely adopted in standards such as 802.11, LTE and 5G NR, has become a fundamental modulation technique in modern wireless communication systems due to its high spectral efficiency and inherent robustness to multipath propagation \cite{1096357, 1090705}. In an OFDM system, a cyclic prefix (CP) is appended to each symbol to mitigate inter-symbol interference (ISI), and the frequency-selective channel is effectively transformed into parallel flat-fading subchannels, thereby simplifying the process of equalization and suppressing the inter-carrier interference (ICI). Despite these advantages, OFDM remains vulnerable to both narrowband and wideband interference that overlap in time and frequency, an event that is common as the spectral environment becomes increasingly congested. Such interference can severely degrade the detection performance, particularly when the interference power is high and its occupied bandwidth substantially overlaps with that of the OFDM signal \cite{1374934, 4099573}. In practical wireless communication systems, forward error correction (FEC) is commonly employed to maintain link reliability by correcting noise-induced symbol and bit errors, especially in noise-limited channels. However, strong overlapping interference can violate the noise assumptions used in soft demodulation and decoding, producing unreliable log-likelihood ratios (LLRs) and limiting the effectiveness of FEC \cite{8299687}. Thus, interference cancellation or interference-aware decoding are still required. \par
Motivated by these challenges, extensive research efforts have been devoted to interference mitigation techniques for OFDM systems. Broadly, interference mitigation can exploit the redundancy and degrees of freedom across the time, frequency and spatial domains, in addition to receiver-side estimation and filtering techniques \cite{10736552, 9405679}. For narrowband interference (NBI) mitigation, frequency-domain cancellation approaches have been widely investigated, since such interference typically occupies only a limited portion of the transmitted bandwidth. In particular, the methods in \cite{1002922,996900, 1512110} distribute coded or spread information across multiple subcarriers to reduce sensitivity to localized interference. In \cite{4960156}, prediction-error filtering is used to localize narrowband interference to a limited set of subcarriers, while bit-interleaved coded modulated (BICM) coding provides the redundancy needed to recover the affected information. Since NBI is typically sparse in the frequency-domain, compressive-sensing-based cancellation methods have also been explored in \cite{5751191, 5683790, 7918877}. In addition, a successive interference cancellation strategy is discussed in \cite{4114231}. The rank-reduced LMMSE-based canceller proposed by \cite{1256751} estimates the interference in the frequency-domain and subtracts the linear minimum mean-square error (LMMSE) estimate from the received signal. In parallel, excision-based filtering methods have been studied, where corrupted subcarriers are nulled prior to demodulation, as analyzed by \cite{1687772}. Similar excision strategies have also been widely reported in the spread-spectrum literature for suppressing narrowband interference \cite{771045, 1094725, 705409}. \par
For interference occupying a substantial portion of the desired-signal bandwidth, mitigation can also exploit redundancy across frequency and space. Frequency-spread coding and interleaving combined with adaptive co-channel interference cancellation for OFDM systems have been investigated in \cite{shibahara2003performance}, while adaptive antenna arrays that exploit spatial diversity for co-channel interference suppression have been studied in \cite{752127}. In \cite{1687750}, the authors further considered an OFDM-based UWB approach that exploits redundant observations obtained through frequency spreading, time/frequency repetition, and multiple receive antennas. Another class of approaches explicitly estimates and reconstructs the interference waveform. Coherent time-domain cancellation (CTC) techniques have been proposed in \cite{ellingson2022coherent, li2025parametric, 11435250}, where the interference is first estimated, reconstructed and subtracted from the received waveform in the time-domain. However, the approaches in \cite{li2025parametric} and \cite{11435250}, while effective for modulated interference, rely on conditions in which either the interference power is significantly higher than that of the desired signal, or the desired signal bandwidth is substantially larger than that of the interference. \par
In contrast to approaches that primarily rely on coding or spreading redundancy in the desired signal, or explicitly estimate and reconstruct the interference, interference mitigation can also exploit inherent structural redundancy in modulated waveforms. In many practical communication, radar, and sonar scenarios—such as interference from unintended users or neighboring cells—the interfering signals exhibit cyclostationarity, meaning their second-order statistics are periodic in time. For cyclostationary signals, certain frequency-shifted versions of the waveform are strongly correlated with the original signal. These inherent correlations can be exploited by a periodic time-varying frequency-shift (FRESH) filter to enhance signal recovery. The FRESH filter is the optimal linear filter for cyclostationary processes, generalizing Wiener’s theory of optimal time-invariant filtering for stationary time series to optimal polyperiodic time-varying filtering for cyclostationary signals \cite{212375}. \par
The authors in \cite{5771071} analyzed the cyclostationarity of OFDM signals with CP (CP-OFDM) and derived a closed-form expression for the FRESH filter coefficients based on the MMSE criterion. Although only additive white Gaussian noise (AWGN) and Rayleigh fading channels were considered in \cite{5771071}, the FRESH filter output demonstrated signal-to-noise ratio (SNR) improvement compared to the input. In \cite{6750418}, the FRESH filter was applied to OFDM signal recovery in power line communication systems, where the noise is strongly periodic and cyclostationary in nature. The presence of co-channel interference, where both the desired and interfering signals are carrier- or pulse-modulated, was studied in \cite{1200960}. Recovery of OFDM signals under interference was further investigated in \cite{carrick2018paramorphic}, where cyclostationarity was deliberately introduced within and across OFDM symbols by repeating subcarrier symbols. \par
Despite its superior performance over the conventional Wiener filter and error-correction-based approaches in interference scenarios, the practical implementation of the FRESH filter can be computationally demanding. Specifically, a FRESH filter consists of multiple filter banks corresponding to the cyclic frequencies of both the desired and interfering signals, and the estimation of second-order statistics for obtaining the filter coefficients at each branch incurs significant complexity when the underlying signal parameters vary over time. Moreover, the filter coefficient generation requires known or training symbols for cross-statistical estimation, and the resulting coefficients may become mismatched as the signal or interference parameters evolve. However, when the structure of the desired OFDM signal is known a priori and the interference parameters can be estimated, closed-form expressions for the second-order statistics of both signals and consequently for the FRESH filter coefficients can be derived. This enables offline pre-computation of the filter coefficients. Moreover, in practical receiver implementations, a certain level of timing synchronization is inevitably required to determine the sampling instants and OFDM symbol boundaries. Therefore, once coarse symbol timing has been achieved, the pre-computed filter coefficients can be applied to the received signal following the channel equalization stage to compensate for channel distortions. \par
\textit{Contributions}: In this paper, we first analyze the second-order statistics of OFDM signals with and without a CP under known OFDM signal parameterization and subcarrier allocation structure. The resulting analytical expressions provide the statistical quantities required to derive the FRESH filter coefficients for both pre- and post-CP-removal configurations. Based on this foundation, we develop a FRESH filtering framework with a compact closed-form matrix expression for the optimal coefficients. For the case of FRESH filtering after CP-removal, the removal of the CP from the desired OFDM signal also truncates the interference within each OFDM block. We account for this effect by modeling the resulting variation in interference symbol timing across OFDM blocks as a random variable and derive the corresponding second-order statistics of the interference signal, which are then combined with those of the desired signal to obtain the FRESH filter coefficients. The resulting closed-form formulation allows the FRESH filter coefficients to be determined offline based on the desired signal structure and interference parameters, while variations in the channel conditions can be subsequently accommodated through equalization. Simulation results validate the effectiveness of the derived theoretical expressions. By exploiting the spectral redundancy of both the desired signal and the interference, the FRESH filter is capable of suppressing the interference, even when the interference power is several orders of magnitude higher than that of the desired signal. Furthermore, we demonstrate that combining the FRESH filter with error-correction techniques achieves superior performance compared to employing either approach individually, at the cost of a slight reduction in the spectral efficiency. This performance gain is particularly pronounced in strong interference scenarios. \par
\textit{Organization}: The remainder of this paper is organized as follows. Section~\ref{sec:II} reviews the fundamentals of cyclostationary signals and the FRESH filter. Section~\ref{sec:III} characterizes the cyclostationary properties of OFDM signals with and without CP insertion. Section~\ref{sec:IV} develops the theoretical time- and frequency-domain FRESH filtering frameworks and discusses their computational complexity and practical implementation. Section~\ref{sec:V} provides numerical results which characterize the performance of the FRESH filter. Subsequently, Section~\ref{sec:VI} concludes the paper. 

\section{Preliminaries}
\label{sec:II}
\subsection{Cyclostationary Signals}
A complex-valued random process \(x(t)\) is said to be wide-sense cyclostationary (WSCS) with fundamental period \(T\) if its second-order statistics are periodic in time. Specifically, the autocorrelation function
\begin{equation}
R_x(t,\tau) = \mathbb{E}\left[ x\left(t+\frac{\tau}{2} \right)x^*\left(t-\frac{\tau}{2} \right)\right]
\end{equation}
satisfies \(R_x(t,\tau) = R_x(t+T, \tau)\). Due to this periodicity, \(R_x(t,\tau)\) admits a Fourier series expansion with respect to \(t\), given by
\begin{equation}
R_x(t,\tau) = \sum_{\alpha} R_x^{\alpha}(\tau) e^{j2\pi \alpha t},
\end{equation}
where \(\alpha\) is the cyclic frequency, and \(R_x^{\alpha}(\tau)\) denotes the cyclic autocorrelation function (CAF) at cyclic frequency \(\alpha\). \par
The CAF characterizes cyclostationary structure in the time-lag domain. Transforming this quantity into the frequency-domain yields the spectral correlation function (SCF), which captures the correlation between spectral components separated by the cyclic frequency. Specifically, the SCF is defined as
\begin{equation}
S_x^\alpha (f) = \int_{-\infty}^\infty R_x^{\alpha} (\tau) e^{-j2\pi f \tau} \ \mathrm{d} \tau.
\label{eq:SCF_def_2}
\end{equation}
For a finite number of samples, the SCF can be estimated by block averaging the spectral correlation: 
\begin{equation}
\begin{split}
S_{x}^\alpha(f) = \frac{1}{N_bN} \sum_{b=0}^{N_b-1} X_N\left(bN, f + \frac{\alpha}{2} \right )X_N^*\left(bN, f - \frac{\alpha}{2} \right),
\label{eq:SCF_emp}
\end{split}
\end{equation}
where \(N\) is the block size, \(N_b\) denotes the number of blocks and \(X_N(bN, f)\) is the \(N\) point DFT of the discrete time signal \(x(n)\) for the \(b\)-th block, i.e.,
\begin{equation}
X_N(bN, f) = \sum_{n=0}^{N-1} x(bN + n)e^{-j2\pi f n}, \quad b = 0, 1, \dots, N_b-1.
\end{equation}

\subsection{FRESH Filter}
It is well known that the optimal linear filter for stationary signals is time-invariant. In contrast, for signals exhibiting cyclostationarity with a single period (or multiple incommensurate periods), the corresponding optimal filter is periodically time-varying, with the same periodicity structure. The linear conjugate-linear (LCL) FRESH filter was developed in \cite{212375} to exploit the SCF and, more generally, the conjugate SCF of cyclostationary signals. It is optimal in the minimum mean-square error (MMSE) sense. The LCL FRESH filter incorporates both the SCF and the conjugate SCF, thereby fully exploiting the second-order cyclostationary statistics. In contrast, the linear FRESH filter relies solely on the SCF. When the conjugate SCF of the received signal is zero, linear filtering is sufficient and the use of LCL processing does not provide additional benefit. For typical modulation schemes such as quadrature phase shift keying (QPSK) and higher-order linearly modulated constellations with circular symmetry, the conjugate cyclic autocorrelation vanishes. Therefore, in this work, we focus exclusively on the linear FRESH filter for OFDM signal processing. \par
\begin{figure}[t!]
\begin{center}
\includegraphics[scale = 0.4]{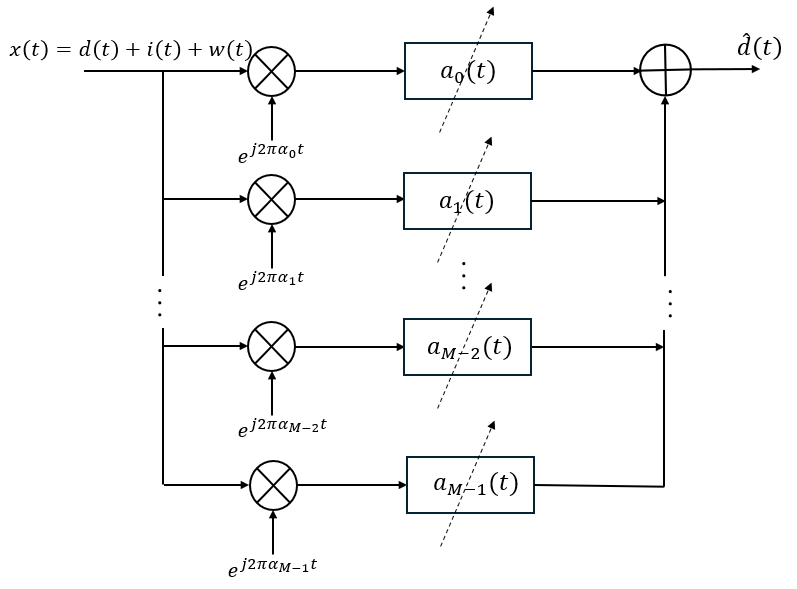}
\caption{Time-domain implementation of a linear FRESH filter. }
\label{fig:FRESH_time}
\end{center}
\end{figure}

Consider the received signal
\begin{equation}
x(t) = d(t) + i(t) + w(t),
\end{equation}
where \(d(t)\) denotes the desired signal, \(i(t)\) denotes the interference signal and \(w(t)\) is the additive white Gaussian noise with zero mean and variance \(\sigma_w^2\). As illustrated in Fig.~\ref{fig:FRESH_time}, the time-domain FRESH filter takes the output of 
\begin{equation}
\hat{d}(t) = \sum_{m = 0}^{M - 1} a_m(t) * x(t)e^{j2\pi \alpha_m t},
\end{equation}
where \(M\) is the number of cyclic frequencies, \(a_m(t)\) and \(\alpha_m\) are the filter coefficient and the cyclic frequency for the \(m\)-th branch, respectively. The set of frequency shifts \(\{\alpha\}\) contains the cyclic frequencies of the desired signal \(\mathcal{A}_d = \{\alpha_{d, 0}, \alpha_{d,1}, \dots\}\), the cyclic frequencies of the interference signal \(\mathcal{A}_i = \{\alpha_{i, 0}, \alpha_{i,1}, \dots\}\) and the pair-wise summation and subtraction between these two sets \((\mathcal{A}_d + \mathcal{A}_i) \cup (\mathcal{A}_d - \mathcal{A}_i)\) due to the additional second-order periodicities created by the linear combination of \(d(t)\) and \(i(t)\). \par

The derivation of the optimal filter coefficients is more conveniently carried out in the frequency-domain, where the FRESH filter output can be expressed as
\begin{equation}
\hat{D}(f) = \sum_{m = 0}^{M - 1} A_m(f) X(f - \alpha_m), 
\label{eq:FRESH_freq}
\end{equation}
where \(\hat{D}(f) = \mathcal{F}\left\{\hat{d}(t) \right\}\), \(X(f) = \mathcal{F}\{x(t) \}\) and \(A_m(f) = \mathcal{F}\{a_m(t)\}\). Here, \(\mathcal{F}\{\cdot\}\) denotes the Fourier transform operation. By taking the derivative of the MSE between the desired signal and its estimate with respect to \(A_m^*(f)\) and setting the result equal to zero, the optimal filter coefficient design equation can be written as \cite{212375}
\begin{equation}
\begin{aligned}
&\sum\limits_{m=0}^{M-1}  S_{x}^{\alpha_p - \alpha_m}\left(f - \frac{\alpha_p + \alpha_m}{2} \right) A_m(f) = S_{dx}^{\alpha_p} \left(f - \frac{\alpha_p}{2} \right), \\
& \qquad p = 0, 1, \dots, M - 1,
\label{eq:FRESH_opt_criterion}
\end{aligned}
\end{equation}
where \(S_{dx}^{\alpha}(f)\) denotes the cross-spectral correlation density between the desired signal \(d\) and the received signal \(x\), defined analogously to (\ref{eq:SCF_emp}) as
\begin{equation}
S_{dx}^{\alpha}(f)
=
\frac{1}{N_bN}
\sum_{b=0}^{N_b-1}
D_N\left(bN, f+\frac{\alpha}{2}\right)
X_N^*\left(bN, f-\frac{\alpha}{2}\right).
\label{eq:cross_SCF_emp}
\end{equation}
A compact matrix representation of (\ref{eq:FRESH_opt_criterion}) for all frequency bins \(f\) can be written as
\begin{equation}
\mathbf{S}_x\mathbf{A} = \mathbf{S}_{dx}
\label{eq:FRESH_opt_freq}
\end{equation}
where 
\begin{equation}
\mathbf{S}_x = 
\begin{bmatrix}
\mathbf{S}_x^{\alpha_0 - \alpha_0} & \mathbf{S}_x^{\alpha_0 - \alpha_1} & \dots & \mathbf{S}_x^{\alpha_0 - \alpha_{M-1}} \\
\mathbf{S}_x^{\alpha_1 - \alpha_0} & \mathbf{S}_x^{\alpha_1 - \alpha_1} & \dots & \mathbf{S}_x^{\alpha_1 - \alpha_{M-1}} \\
\vdots & \vdots & \ddots & \vdots \\
\mathbf{S}_x^{\alpha_{M-1} - \alpha_0} & \mathbf{S}_x^{\alpha_{M-1} - \alpha_1} & \dots & \mathbf{S}_x^{\alpha_{M-1} - \alpha_{M-1}} \\
\end{bmatrix}
\end{equation}
is the SCF matrix where each submatrix \(\mathbf{S}_x^{\alpha_p - \alpha_m}\) is a diagonal matrix with \(S_x^{\alpha_p - \alpha_m}\left( f_i - \frac{\alpha_p + \alpha_m}{2} \right)\) at the \((i,i)\)-th entry. On the other hand, \(\mathbf{S}_{dx}\) is the cross-SCF vector between the desired signal and the received signal, and can be written as
\begin{equation}
\mathbf{S}_{dx} = 
\begin{bmatrix}
\left(\mathbf{S}_{dx}^{\alpha_0}\right)^T & \left(\mathbf{S}_{dx}^{\alpha_1}\right)^T & \dots & \left(\mathbf{S}_{dx}^{\alpha_{M-1}}\right)^T
\end{bmatrix}^T,
\end{equation}
where the \(i\)-th entry of the subvector \(\mathbf{S}_{dx}^{\alpha_p}\) is \(S_{dx}^{\alpha_p} \left(f_i - \frac{\alpha_p}{2} \right)\). Furthermore, 
\begin{equation}
\mathbf{A} = 
\begin{bmatrix}
\mathbf{A}_0^T & \mathbf{A}_1^T & \dots \mathbf{A}_{M-1}^T
\end{bmatrix}^T,
\end{equation}
where the \(i\)-th entry of the subvector \(\mathbf{A}_p\) is \(A_p(f_i)\). \par
Similarly, by taking the inverse Fourier transform of (\ref{eq:FRESH_opt_freq}) along the frequency axis \(f\), we can obtain a linear system for the time-domain filter coefficient, which can be written as
\begin{equation}
\mathbf{R}_x\mathbf{a} = \mathbf{R}_{dx},
\label{eq:FRESH_opt_time}
\end{equation}
where 
\begin{equation}
\mathbf{R}_x =
\begin{bmatrix}
\mathbf{R}_x^{\alpha_0 - \alpha_0} & \mathbf{R}_x^{\alpha_0 - \alpha_1} & \dots & \mathbf{R}_x^{\alpha_0 - \alpha_{M-1}} \\
\mathbf{R}_x^{\alpha_1 - \alpha_0} & \mathbf{R}_x^{\alpha_1 - \alpha_1} & \dots & \mathbf{R}_x^{\alpha_1 - \alpha_{M-1}} \\
\vdots & \vdots & \ddots & \vdots \\
\mathbf{R}_x^{\alpha_{M-1} - \alpha_0} & \mathbf{R}_x^{\alpha_{M-1} - \alpha_1} & \dots & \mathbf{R}_x^{\alpha_{M-1} - \alpha_{M-1}} \\
\end{bmatrix},
\end{equation}
is the CAF matrix where the \((i,j)\)-th entry of each submatrix \(\mathbf{R}_x^{\alpha_p - \alpha_m}\) is \(R_x^{\alpha_p - \alpha_m}(\tau_{i-j})\). Similarly, 
\begin{equation}
\mathbf{R}_{dx} = 
\begin{bmatrix}
\left(\mathbf{R}_{dx}^{\alpha_0}\right)^T & \left(\mathbf{R}_{dx}^{\alpha_1}\right)^T & \dots & \left(\mathbf{R}_{dx}^{\alpha_{M-1}}\right)^T
\end{bmatrix}^T,
\end{equation}
is the cross-CAF vector where the \(i\)-th entry of the subvector \(\mathbf{R}_{dx}^{\alpha_{m}}\) is \(R_{dx}^{\alpha_p}(\tau_i)\). Furthermore, 
\begin{equation}
\mathbf{a} = 
\begin{bmatrix}
\mathbf{a}_0^T & \mathbf{a}_1^T & \dots \mathbf{a}_{M-1}^T
\end{bmatrix}^T,
\end{equation}
where the \(i\)-th entry of the subvector \(\mathbf{a}_p\) is \(a_p(\tau_i)\). 

\section{CYCLOSTATIONARITY OF OFDM SIGNAL}
\label{sec:III}
In this section, we first analyze the CAF of the CP-OFDM which serves for the time-domain FRESH filter placed before CP-removal. We then analyze the SCF of the CP-free OFDM signal which serves for the frequency-domain FRESH filter placed after CP-removal. Although the SCF of the CP-OFDM signal can be obtained directly from its CAF, analyzing the CP-removed signal leads to a much simpler frequency-domain representation and a lower computational complexity while yielding the required statistics for frequency-domain FRESH filtering.

\subsection{CAF of OFDM signal}
In general, a transmitted OFDM signal can be written as
\begin{equation}
\begin{split}
d(n) &= \frac{1}{\sqrt{N}} \sum_{l = -\infty}^\infty \sum_{k = 0}^{N-1} X_{l, k} e^{j2\pi (n - lN_s - N_{\mathrm{cp}}) \frac{k}{N}} g(n - lN_s),
\end{split}
\end{equation}
where \(N\) is the total number of subcarriers, \(N_{\mathrm{cp}}\) is the cyclic prefix length and \(N_s = N + N_{\mathrm{cp}}\). Pulse shape \(g(n)\) is an \(N_s\)-long rectangular window. The frequency-domain symbol for the \(k\)-th subcarrier and \(l\)-th OFDM symbol \(X_{l, k}\) is obtained by mapping the information symbol onto subcarriers. Specifically, define the information symbols for the \(l\)-th OFDM symbol as \(\mathbf{s}_l = [s_{l, 1}, \dots, s_{l, K}]^T\in \mathbb{C}^{K}\). These symbols are mapped to the \(N\) subcarriers through a precoding matrix \(\mathbf{P} \in \mathbb{C}^{N \times K}\). If the \(k\)-th subcarrier is nulled, then the \(k\)-th row of \(\mathbf{P}\) is zero; if the \(p\)-th symbol is assigned to the \(k\)-th subcarrier, then \(P_{k, p} = 1\). The resulting frequency-domain OFDM symbol vector is \(\mathbf{X}_l = \mathbf{Ps}_l\), where \(X_{l, k} = \sum_{p=0}^{K-1}P_{k,p}s_{l, p}\). \par
Now, the time-domain autocorrelation function of the OFDM signal can be expressed as
\begin{equation}
\begin{split}
R_d(n, \tau) &= \mathbb{E}\left[d\left(n + \frac{\tau}{2} \right)d^*\left(n - \frac{\tau}{2} \right) \right], \\
&= \frac{1}{N} \sum_{l = -\infty}^\infty \sum_{l' = -\infty}^\infty \sum_{k = 0}^{N-1} \sum_{k' = 0}^{N-1} \mathbb{E}[X_{l, k} X_{l', k'}^*]  \\
&\quad \ e^{j2\pi (n + \tau/2 - lN_s - N_{\mathrm{cp}}) \frac{k}{N}} e^{-j2\pi (n  - \tau/2 - l'N_s - N_{\mathrm{cp}}) \frac{k'}{N}} \\
& \quad \ g(n + \tau/2 -lN_s)  g(n - \tau/2 -l'N_s),\\
&= \frac{1}{N}  \sum_{l = -\infty}^\infty \sum_{k = 0}^{N-1} \sum_{k' = 0}^{N-1} \left[\mathbf{PP}^H \right]_{k, k'} \\
&\quad \ e^{j\frac{2\pi}{N} [(k - k')(n - lN_s -N_{\mathrm{cp}}) + \tau/2(k + k')]} \\
&\quad \ g(n + \tau/2 -lN_s)  g(n - \tau/2 -lN_s),
\end{split}
\end{equation}
where \(\left[ \mathbf{PP}^H \right]_{k, k'}\) denotes the \((k,k')\)-th entry of the matrix \(\mathbf{PP}^H\), which represents the correlation between the \(k\)-th and \(k'\)-th subcarrier and we have used the assumption that the frequency-domain symbols are independent across different OFDM symbols. \par
Since \(R_d(n, \tau)\) is an \(N_s\)-length periodic function with respect to \(n\), the discrete Fourier series of the autocorrelation function is the CAF, which can be written as
\begin{equation}
\begin{split}
R_d^\alpha(\tau) &= \frac{1}{N_s} \sum_{n=0}^{N_s-1} R_d(n, \tau) e^{-j2\pi \alpha n}, \\
&= \frac{1}{NN_s} \sum_{n=0}^{N_s-1} \sum_{l = -\infty}^\infty \sum_{k = 0}^{N-1} \sum_{k' = 0}^{N-1} \left[\mathbf{PP}^H \right]_{k, k'} \\
&\quad \ e^{j\frac{2\pi}{N} [(k - k')(n - lN_s -N_{\mathrm{cp}}) + \tau/2(k + k')]} \\
&\quad \ g(n + \tau/2 -lN_s)  g(n - \tau/2 -lN_s) e^{-j2\pi \alpha n},\\ 
&= \frac{1}{NN_s} \sum_{k = 0}^{N-1} \sum_{k' = 0}^{N-1} \left[\mathbf{PP}^H \right]_{k, k'} e^{-j2\pi (k - k')\frac{N_{\mathrm{cp}}}{N}} e^{j\pi \tau \frac{k + k'}{N}} \\
&\qquad \qquad \qquad \qquad  \ R_g^{\alpha - \frac{k - k'}{N} } (\tau),
\label{eq:CAF_CP_OFDM}
\end{split}
\end{equation}
where 
\begin{equation}
R_g^{\alpha}(\tau) = \sum_{n = -\infty}^\infty g\left(n + \frac{\tau}{2} \right) g\left(n - \frac{\tau}{2} \right) e^{-j2\pi \alpha n}.
\end{equation}\par
For the case when all subcarriers are independent, \(\mathbf{PP}^H\) is an identity matrix so that only the pairs \((k,k')\) with \(k = k'\) in (\ref{eq:CAF_CP_OFDM}) survive. In this case, the CAF reduces to the expression in \cite{5771071}:
\begin{equation}
R_d^\alpha(\tau) = \frac{1}{NN_s} \sum_{k = 0}^{N-1} e^{j2\pi \tau \frac{k}{N}} R_g^\alpha(\tau).
\end{equation}

\begin{figure}[t!]
\begin{center}
\includegraphics[scale = 0.17]{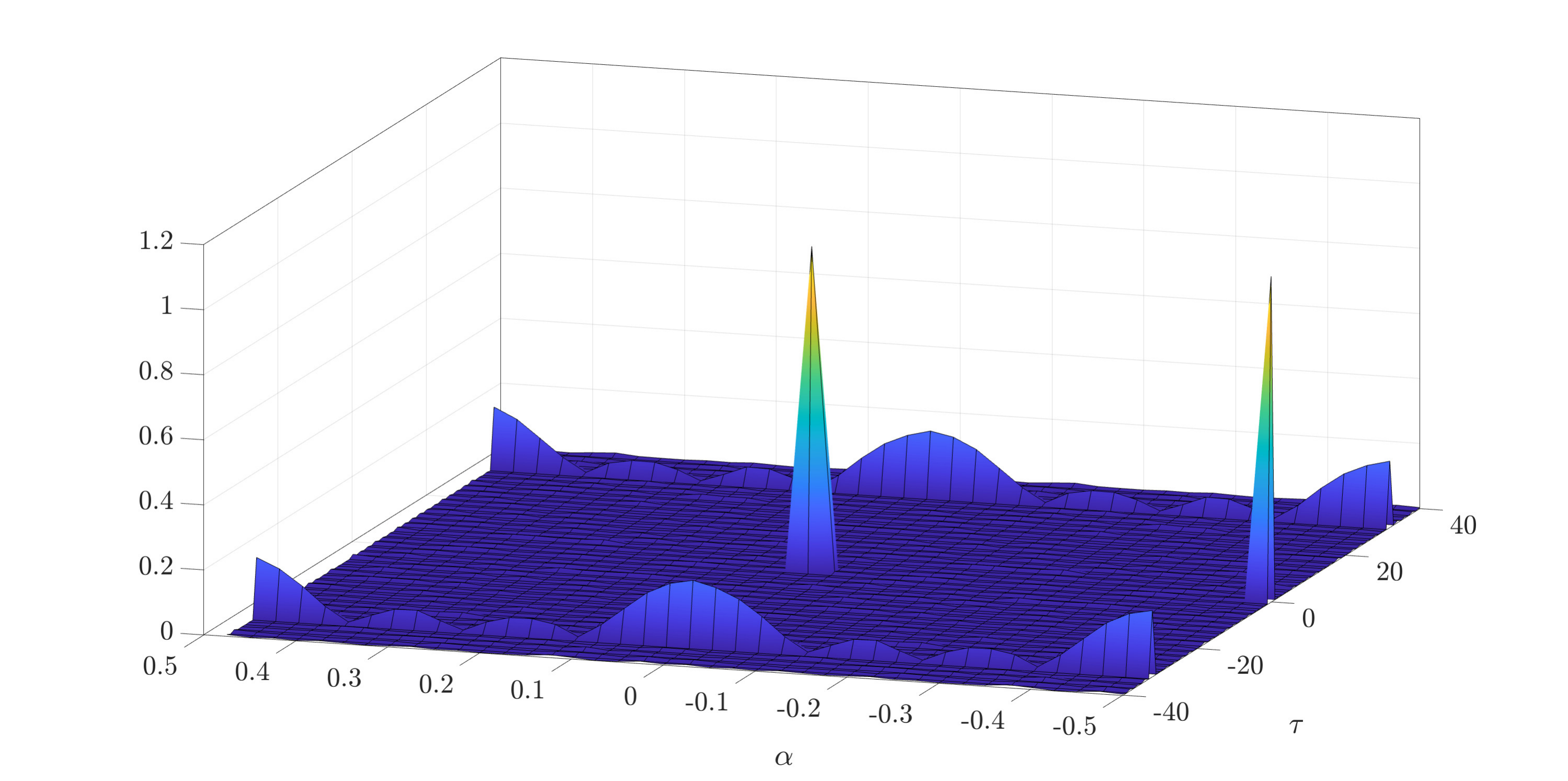}
\caption{CAF of the OFDM signal with frequency-domain symbol repetition factor of two, where \(N=32\) and \(N_{\mathrm{cp}}=8\).  }
\label{fig:CAF_redun2}
\end{center}
\end{figure}
On the other hand, when the subcarriers are not mutually independent, \(\mathbf{PP}^H\) contains nonzero off-diagonal elements corresponding to the correlation between the dependent subcarriers. Fig.~\ref{fig:CAF_redun2} illustrates the resulting CAF when correlation is introduced into the frequency-domain through symbol repetition by a factor of two, where the second half of the subcarriers contains a copy of the first half. The dominant cyclic components occur at \(\alpha=0\) and \(\alpha=\frac{1}{2}\), which are introduced by the frequency-domain repetition. In addition, weaker components appear at time lags \(\tau=\pm N\), which are induced by the CP.

\subsection{SCF of OFDM signal}
As shown in (\ref{eq:SCF_def_2}), the SCF can be written as the Fourier transform of the CAF in (\ref{eq:CAF_CP_OFDM}) along the delay variable \(\tau\), which results in an SCF for the CP-OFDM that can be used for obtaining the optimal filter coefficient of the FRESH filter before CP-removal. However, since CP removal can be readily performed once symbol timing has been established, the SCF analysis can be simplified by focusing on the CP-removed OFDM signal, as the frequency-domain representation consists of linearly modulated subcarrier symbols, thereby significantly reducing the computational complexity of the FRESH filter placed after CP removal. As indicated by (\ref{eq:SCF_emp}), if the block size is equal to the number of subcarriers \(N\), the SCF across all frequency bins can be conveniently expressed through element-wise multiplication between two shifted versions of the frequency-domain OFDM symbols. Specifically, under the assumption of cyclo-ergodicity, the SCF for the CP-removed OFDM signal can be expressed as
\begin{equation}
\begin{split}
S_d^\alpha(f) &= \mathbb{E}\left[D_N\left(f + \frac{\alpha}{2} \right) D_N^*\left(f - \frac{\alpha}{2} \right) \right], \\
&= \mathbb{E}\left[ [\mathbf{Ps}]_{f + \alpha/2} [\mathbf{Ps}]_{f - \alpha/2}^* \right], \\
&= \left[ \mathbf{PP}^H \right]_{f + \alpha/2, f - \alpha/2},
\label{eq:SCF_OFDM}
\end{split}
\end{equation}
where \([\mathbf{Ps}]_{k}\) denotes the \(k\)-th entry of the vector \(\mathbf{Ps}\) and the indices are taken modulo \(N\). \par
To facilitate the theoretical expression of the SCF used for obtaining the optimal filter coefficients in the FRESH filter~(\ref{eq:FRESH_opt_freq}), the SCF subjected to the cyclic frequencies \(\alpha_m\) and \(\alpha_p\) can be written as
\begin{equation}
\begin{split}
S_d^{\alpha_p - \alpha_m}\left(f - \frac{\alpha_p + \alpha_m}{2} \right) &= \mathbb{E}\left[D_N(f - \alpha_m)D_N^*(f - \alpha_p) \right], \\
&= \left[ \mathbf{PP}^H \right]_{f - \alpha_m, f - \alpha_p}.
\label{eq:SCF_OFDM_FRESH}
\end{split}
\end{equation} 
It is more convenient to express the SCF evaluated over all discrete frequency bins, which can be written as
\begin{equation}
\begin{split}
\begin{bmatrix}
S_d^{\alpha_p - \alpha_m}(f_0) \\
S_d^{\alpha_p - \alpha_m}(f_1) \\
\vdots \\
S_d^{\alpha_p - \alpha_m}(f_{N-1}) 
\end{bmatrix}
&= \mathbb{E}[\mathbf{C}_{\alpha_m}\mathbf{Ps} \odot \mathbf{C}_{\alpha_p}\mathbf{Ps}], \\
&= \mathrm{diag}\left(\mathbf{C}_{\alpha_m} \mathbf{PP}^H  \mathbf{C}_{\alpha_p}^H \right),
\label{eq:SCF_vec_theory}
\end{split}
\end{equation}
where \(\odot\) represents the element-wise product and \(\mathbf{C}_\alpha\) is the permutation matrix which circularly shifts the vector by \(\alpha N\) frequency bins. \par

\section{FRESH FILTER USING THEORETICAL EXPRESSION}
\label{sec:IV}
In this section, we present the framework of the FRESH filter utilizing the theoretical second-order statistics of the OFDM symbol derived in the previous section. In conventional estimation-based FRESH filtering, the required CAF/SCF for obtaining the filter coefficients are estimated from signal realizations, requiring the statistics to be re-estimated for each realization. In particular, estimating the cross-CAF/SCF requires a known or training sequence of the desired signal. In contrast, the analytical framework developed here enables the FRESH filter coefficients to be determined directly from the theoretical second-order statistics once the signal structure and the associated parameters are known. 

\subsection{Time-domain FRESH filter}
Under the assumption that the desired signal, interference, and noise are mutually independent, the cross-CAF for each delay \(\tau\) in \(\mathbf{R}_{dx}^{\alpha_p}\) on the right side of (\ref{eq:FRESH_opt_time}) reduces to the CAF of the desired signal \(R_{d}^{\alpha_p}(\tau_i)\), which can be obtained through (\ref{eq:CAF_CP_OFDM}) for a given precoding matrix \(\mathbf{P}\). On the other hand, the CAF for each delay \(\tau\) in \(\mathbf{R}_x^{\alpha_p - \alpha_m}\) can be written as
\begin{equation}
\begin{split}
R_x^{\alpha_p - \alpha_m}(\tau_{i-j}) &= R_d^{\alpha_p - \alpha_m}(\tau_{i-j}) + R_i^{\alpha_p - \alpha_m}(\tau_{i-j}) \\
&\quad \ + R_w^{\alpha_p - \alpha_m}(\tau_{i-j}),
\end{split}
\end{equation}
where \(R_i^{\alpha}(\tau)\) and \(R_w^{\alpha}(\tau)\) denote the CAF for the interference signal and noise, respectively. \par
The first term can be obtained through (\ref{eq:CAF_CP_OFDM}). On the other hand, the CAF for the interference \(R_i^{\alpha}(\tau)\) depends on the interference type. If the interference is also OFDM signal, the SCF can be expressed in the same form as (\ref{eq:CAF_CP_OFDM}). In contrast, the CAF of various signal types is well established once the interference signal parameters are estimated, as discussed in Chapter 12 of \cite{gardner1989statistical}. For a single-carrier interference signal, these parameters include the symbol duration, frequency offset, and pulse-shaping parameters. \par
Lastly, since the Gaussian noise samples \(w(n)\) are independent to each other, the CAF of the noise can be written as
\begin{equation}
R_{w}^{\alpha_m - \alpha_p}(\tau_{i-j}) = \sigma_w^2 \delta(\alpha_p - \alpha_m) \delta(i - j).
\end{equation} 
Thus, given the desired-signal and interference parameters, \(\mathbf{R}_x\) and \(\mathbf{R}_{dx}\) in (\ref{eq:FRESH_opt_time}) can be analytically determined, allowing the time-domain FRESH filter coefficients to be obtained offline.

\subsection{Frequency-domain FRESH filter}
Under the assumption that the desired signal, interference, and noise are mutually independent, the cross-SCF for each frequency bin \(f\) in \(\mathbf{S}_{dx}^{\alpha_p}\) on the right side of (\ref{eq:FRESH_opt_freq}) reduces to the SCF of the desired signal \(S_{d}^{\alpha_p} \left(f - \frac{\alpha_p}{2} \right)\). As indicated by (\ref{eq:SCF_vec_theory}), by setting the \(\alpha_m = 0\), the cross-SCF vector can be written as
\begin{equation}
\mathbf{S}_{d}^{\alpha_p} = \mathrm{diag}\left(\mathbf{PP}^H  \mathbf{C}_{\alpha_p}^H \right).
\end{equation}
On the other hand, the SCF for each frequency bin in \(\mathbf{S}_x^{\alpha_p - \alpha_m}\) on the left side of (\ref{eq:FRESH_opt_freq}) can be written as 
\begin{equation}
\begin{split}
S_{x}^{\alpha_p - \alpha_m}\left(f - \frac{\alpha_p + \alpha_m}{2} \right) &= S_{d}^{\alpha_p - \alpha_m}\left(f - \frac{\alpha_p + \alpha_m}{2} \right) \\ 
& \quad + S_{i}^{\alpha_p - \alpha_m}\left(f - \frac{\alpha_p + \alpha_m}{2} \right) \\
& \quad + S_{w}^{\alpha_p - \alpha_m}\left(f - \frac{\alpha_p + \alpha_m}{2} \right).
\label{eq:SCF_d_i_w}
\end{split}
\end{equation}
The first term can be obtained through (\ref{eq:SCF_OFDM_FRESH}). \par
The SCF of the interference \(S_i^{\alpha_p-\alpha_m}(f)\) in (\ref{eq:SCF_d_i_w}) requires additional consideration in the frequency-domain implementation. Similar to the time-domain approach, it depends on the type of interference signal. Moreover, it is affected by the OFDM block length \(N\) due to the block-wise structure of the signal, as well as the CP length \(N_{\mathrm{cp}}\). Specifically, when the CP is removed, the first \(N_{\mathrm{cp}}\) samples of the interference within each OFDM block are discarded, which may alter its cyclostationary characteristics. To obtain the truncated block-wise SCF of the interference signal, we first derive the SCF for the block-wise interference signal with block length equal to \(N\) without considering the impact of the truncation due to CP removal. 
Specifically. 
\begin{equation}
S_{i}^{\alpha_p - \alpha_m}\left(f - \frac{\alpha_p + \alpha_m}{2} \right) = \mathbb{E}[I_N(f - \alpha_m)I_N^*(f - \alpha_p)].
\end{equation}
Here, \(I_N(f)\) denotes the frequency-domain representation of the interference obtained by applying the Fourier transform over a finite observation interval of \(N\) samples. Equivalently, this finite-length observation can be viewed as multiplying the time-domain interference \(i(t)\) by a rectangular window, which corresponds to convolving its spectrum with a sinc function in the frequency-domain. Let \(I(f)\) denote the spectrum of the interference signal over an infinite observation interval. The frequency-domain representation of the \(N\)-sample observation can then be expressed as
\begin{equation}
I_N(f) = \int_{-\infty}^\infty I(\nu) G(f - \nu) \ \mathrm{d}\nu,
\end{equation}
where \(G(f) = \sum_{n=0}^{N-1} g(n)e^{-j2\pi f n} =  e^{-j\pi f (N-1)} \frac{\sin(\pi N f)}{\sin(\pi f)}\) is the frequency-domain rectangular window corresponding to \(g(n)\). Substituting this expression into the block-wise SCF yields
\begin{equation}
\begin{split}
&S_{i}^{\alpha_p - \alpha_m}\left(f - \frac{\alpha_p + \alpha_m}{2} \right) = \mathbb{E}\bigg[ \int_{-\infty}^\infty I(\nu) G(f - \alpha_m - \nu) \ \mathrm{d}\nu \\
& \qquad \qquad \qquad \qquad \qquad \quad \cdot\int_{-\infty}^\infty I^*(\mu) G^*(f - \alpha_p - \mu) \ \mathrm{d}\mu  \bigg], \\
&= \int_{-\infty}^\infty \int_{-\infty}^\infty \underbrace{\mathbb{E}\left[ I\left(\xi + \frac{\gamma}{2} \right) I^*\left(\xi - \frac{\gamma}{2} \right)\right]}_{\tilde{S}_i^\gamma(\xi)} \\
& \quad \ \cdot \underbrace{G\left( f - \alpha_m - \xi - \frac{\gamma}{2} \right) G^*\left( f - \alpha_p - \xi + \frac{\gamma}{2} \right)}_{\Phi_{p,m}^\gamma(f, \xi)} \ \mathrm{d}\xi \ \mathrm{d}\gamma, \\
\label{eq:SCF_windowed}
\end{split}
\end{equation}
where we have set \(\nu = \xi + \frac{\gamma}{2}\), \(\mu = \xi - \frac{\gamma}{2}\). Under the assumption that the interference is second-order cyclo-ergodic, the quantity
\begin{equation}
\tilde{S}_i^\gamma(\xi) = \mathbb{E}\left[ I\left( \xi + \frac{\gamma}{2}\right) I^*\left( \xi - \frac{\gamma}{2}\right) \right]
\label{eq:SCF_ideal}
\end{equation}
corresponds to the true SCF defined in (\ref{eq:SCF_def_2}). Equation (\ref{eq:SCF_windowed}) therefore shows that the expected value of the block-averaged SCF estimator corresponds to a two-dimensional convolution of the true interference SCF with the frequency response of a rectangular window \(G(f)\). \par
For most practical signals, the SCF is nonzero only at a discrete set of cyclic frequencies. Therefore, the integral over \(\gamma\) can be replaced by the summation over the possible set of the cyclic frequencies \(\mathcal{A}_i\), yielding
\begin{equation}
\begin{split}
S_{i}^{\alpha_p - \alpha_m}\left(f - \frac{\alpha_p + \alpha_m}{2} \right) = \sum_{\gamma \in \mathcal{A}_i} \int_{-\infty}^\infty \tilde{S}_i^\gamma(\xi) \Phi_{p,m}^\gamma(f, \xi) \ \mathrm{d}\xi.
\end{split}
\end{equation}
Similar to the CAF of the interference for the time-domain FRESH filter, the SCF of various signal types is well established once their signal parameters are known \cite{gardner1989statistical}. \par
Now, when removing the CP of the OFDM signal, the first \(N_{\mathrm{cp}}\) samples of the interference are also removed within each \(N_s\)-long block. Assuming that the symbol duration of the interference signal and the number of subcarriers \(N\) are independent, the symbol timing of the interference signal may vary for each block. Since timing offset in time-domain corresponds to phase shift in frequency-domain, the SCF evaluated through (\ref{eq:SCF_windowed}) and (\ref{eq:SCF_ideal}) also admits a frequency shift, i.e., 
\begin{equation}
\begin{split}
&\mathbb{E}\left[\mathcal{F}\{i(n - n_0) e^{-j\pi \gamma \xi } \} \mathcal{F}\{i^*(n - n_0) e^{j\pi \gamma \xi }\} \right] \\
&= \mathbb{E}\left[ I\left(\xi + \frac{\gamma}{2}  \right) e^{-j2\pi (\xi + \gamma/2) n_0 }  I^*\left(\xi + \frac{\gamma}{2}  \right) e^{j2\pi (\xi - \gamma/2) n_0 } \right], \\
&= \tilde{S}_i^\gamma(\xi) e^{-j2\pi \gamma n_0},
\end{split}
\end{equation}
where \(n_0\) is the symbol timing offset. Therefore, the optimal filtering strategy is filtering each block with the filter that is obtained for the corresponding interference symbol timing for each block. However, this requires explicit knowledge of the symbol timing for each block in order to compute the corresponding filter coefficients. In contrast, given that the number of subcarriers \(N\) and CP length \(N_{\mathrm{cp}}\) are independent of the interference symbol duration, we can assume that symbol timing offset for each block follows a uniform distribution within the interference symbol duration, so that the truncated windowed SCF across all blocks can be written as
\begin{equation}
\begin{split}
\hat{S}_{i}^{\alpha_p - \alpha_m}\left(f - \frac{\alpha_p + \alpha_m}{2} \right) &= \sum_{n_0 = 0}^{N_0 - 1}  S_{i}^{\alpha_p - \alpha_m}\left(f - \frac{\alpha_p + \alpha_m}{2} \right) \\
&\qquad \quad \  e^{-j2\pi \left( f - \frac{\alpha_p + \alpha_m}{2} \right) n_0},
\label{eq:SCF_interf_final}
\end{split}
\end{equation}
where \(N_0\) denotes the symbol duration of the interference signal which can be estimated through the cyclic spectrum of the received signal. Although the filter coefficients obtained using the interference SCF evaluated by taking the expectation with respect to the symbol timing offset are sub-optimal, we will show in Section~\ref{sec:V} that the sub-optimal filter coefficients demonstrate similar estimation accuracy of the desired signal. \par
Lastly, since the Gaussian noise samples \(w(n)\) are independent to each other, the estimated SCF of the noise can be written as
\begin{equation}
\hat{S}_w^{\alpha_p - \alpha_m}(f) = \sigma_w^2 \delta(\alpha_p - \alpha_m).
\end{equation}
Thus, the same desired-signal and interference parameters required for the time-domain FRESH filter provide the underlying signal statistics for the frequency-domain formulation. These results provide the SCFs required to construct \(\mathbf{S}_x\) and \(\mathbf{S}_{dx}\) in (\ref{eq:FRESH_opt_freq}), from which the frequency-domain FRESH filter coefficients can be obtained.

\subsection{Complexity Analysis}
\label{sec:IVC}
This subsection compares the computational complexity of the time- and frequency-domain FRESH filter implementations under both empirical and analytical coefficient generation. The time-domain implementation requires convolution for each cyclic-frequency branch. For the frequency-domain implementation, two filtering realizations may be considered. Linear convolution can be implemented using overlap-save processing, which reproduces the streamwise time-domain filtering operation but requires additional buffering, and an extended FFT/IFFT processing stage. In contrast, the block-wise circular implementation operates directly on each CP-removed \(N\)-point FFT output already available in a conventional OFDM receiver, allowing the FRESH filter to be implemented primarily through frequency-domain shifts and multiplications. Therefore, the circular implementation is adopted in the following complexity analysis. As shown later in Section~\ref{sec:VA}, this choice introduces only a modest performance loss while preserving the low-complexity structure of the frequency-domain implementation. The complexity analysis further distinguishes between offline coefficient generation and online processing: the empirical approach requires the second-order statistics and filter coefficients to be obtained for each realization, while the analytical approach incurs a one-time coefficient-generation cost that can be reused when the underlying signal and interference parameters remain unchanged. \par
For the time-domain empirical-based implementation, let \(L\) denote the number of filter taps per cyclic frequency, \(M\) the number of cyclic frequencies, \(N_t\) the training-signal length, and \(N_{\mathrm{rx}}\) the received-signal length. The empirical approach estimates \(ML\) cross-CAF quantities and \(M^2(2L-1)\) CAF quantities, resulting in an overall statistical-estimation complexity of \(\mathcal{O}(N_tM^2L)\). Solving the \(LM\times LM\) linear system requires \(\mathcal{O}(L^3M^3)\), while applying the \(M\) \(L\)-tap FRESH branches to the received signal requires \(\mathcal{O}(N_{\mathrm{rx}}LM)\). Thus,
\begin{equation}
C_{\mathrm{TD,\, empirical}} = \mathcal{O}\left(N_tM^2L+L^3M^3+N_{\mathrm{rx}}LM\right).
\end{equation}

In contrast, the analytical time-domain approach avoids realization-dependent statistical estimation by constructing the required CAFs directly from the known desired-signal structure and estimated interference parameters. For the initial coefficient generation, the analytical OFDM CAF in (\ref{eq:CAF_CP_OFDM}) requires a double summation over the \(N\) subcarriers for each of the \(ML\) cyclic-frequency/lag pairs, resulting in a complexity of \(\mathcal{O}(MN^2L)\). Assuming that the interference parameters have been obtained, the theoretical CAF of a pulse-shaped single-carrier interferer can be directly evaluated for the required \(ML\) pairs with complexity \(\mathcal{O}(MLN_0)\), where \(N_0\) denotes the interference symbol duration. Including the solution of the \(LM\times LM\) linear system, the initial coefficient-generation complexity is
\begin{equation}
C_{\mathrm{TD,\, analytical}}^{\mathrm{offline}} = \mathcal{O}\left(MN^2L+MLN_0+L^3M^3\right).
\end{equation}
Since the CAF of the desired OFDM signal can be pre-computed from the known precoding matrix, subsequent coefficient updates only require updating the interference statistics when the interference parameters vary. The interference CAF is then recalculated, followed by reconstruction of \(\mathbf{R}_x\) and the solution of the updated linear system. Therefore, the coefficient-update complexity is
\begin{equation}
C_{\mathrm{TD,\, analytical}}^{\mathrm{update}} = \mathcal{O}\left(MLN_0 + L^3M^3\right).
\end{equation}
Afterwards, only the filtering operation is required for each new realization:
\begin{equation}
C_{\mathrm{TD,\, analytical}}^{\mathrm{online}} = \mathcal{O}(N_{\mathrm{rx}}LM).
\end{equation}

For the frequency-domain implementation, let \(K_t=N_t/(N+N_{\mathrm{cp}})\) and \(K_r=N_{\mathrm{rx}}/(N+N_{\mathrm{cp}})\) denote the numbers of training and received OFDM symbols, respectively. The empirical cross-SCF and received-signal SCF require \(\mathcal{O}(K_tMN)\) and \(\mathcal{O}(K_tM^2N)\), respectively. Owing to the block-diagonal structure of \(\mathbf{S}_x\), the \(MN\times MN\) system can be decomposed into \(N\) independent \(M\times M\) systems, resulting in a complexity of \(\mathcal{O}(NM^3)\). The resulting filter is applied with \(\mathcal{O}(K_rMN)\) complexity. Hence,
\begin{equation}
C_{\mathrm{FD,\, empirical}} = \mathcal{O}\left(K_tM^2N+NM^3+K_rMN\right).
\end{equation}

Similar to the time-domain case, the analytical frequency-domain approach avoids realization-dependent estimation of the second-order statistics. The SCF of the desired-signal can be pre-computed directly from the known precoding matrix, while the interference SCF in (\ref{eq:SCF_interf_final}) is obtained from the estimated interference parameters. Specifically, the interference SCF is evaluated by averaging the \(N\)-point FFT-based circular-convolution results over \(N_0\) possible symbol timings for each of the \(M^2\) cyclic frequency pairs, resulting in a complexity of \(\mathcal{O}(M^2N_0N\log N)\). The coefficient generation therefore requires
\begin{equation}
C_{\mathrm{FD,\, analytical}}^{\mathrm{offline}} = \mathcal{O}\left(M^2N_0N\log N+NM^3\right).
\end{equation}
As in the time-domain case, the desired-signal SCF can be reused as long as the precoding structure remains unchanged. On the other hand, since the interference parameters such as the symbol rate and pulse-shaping parameters typically vary slowly, updates may mainly be required for faster-varying parameters such as the carrier-frequency offset. In this case, the interference SCF can be updated by shifting the previously computed SCF, so that the dominant coefficient-update complexity becomes
\begin{equation}
C_{\mathrm{FD,\, analytical}}^{\mathrm{update}} = \mathcal{O}\left(NM^3\right).
\end{equation}
Once the updated coefficients are obtained, applying the frequency-domain FRESH filter requires
\begin{equation}
C_{\mathrm{FD,\, analytical}}^{\mathrm{online}} = \mathcal{O}(K_rMN).
\end{equation}

The resulting complexity comparison is summarized in Table~\ref{tab:complexity_comparison}.

\begin{table}[t]
\centering
\caption{Computational complexity of the four FRESH filtering approaches.}
\label{tab:complexity_comparison}
\renewcommand{\arraystretch}{1.25}
\setlength{\tabcolsep}{3pt}
\begin{tabular}{lcc}
\hline
\textbf{Approach} & \textbf{Offline/Update} & \textbf{Online} \\
\hline

TD empirical
& --
& \(\mathcal{O}\big(N_tM^2L+L^3M^3\)
\\[-1mm]
& & \(+N_{\mathrm{rx}}LM\big)\)
\\[1mm]

TD analytical
& \(\mathcal{O}\big(MN^2L+MLN_0+L^3M^3\big)/\)
\\[-1mm]
& \(\mathcal{O}\big(MLN_0+L^3M^3\big)\) 
& \(\mathcal{O}(N_{\mathrm{rx}}LM)\)
\\[1mm]

FD empirical
& --
& \(\mathcal{O}\big(K_tM^2N+NM^3\)
\\[-1mm]
& & \(+K_rMN\big)\)
\\[1mm]

FD analytical
& \(\mathcal{O}\big(M^2N_0N\log N+NM^3\big)/\)
\\[-1mm]
& \(\mathcal{O}\big(NM^3\big)\) 
& \(\mathcal{O}(K_rMN)\)
\\
\hline
\end{tabular}
\end{table}

\begin{figure}[t!]
\begin{center}
\includegraphics[scale = 0.15]{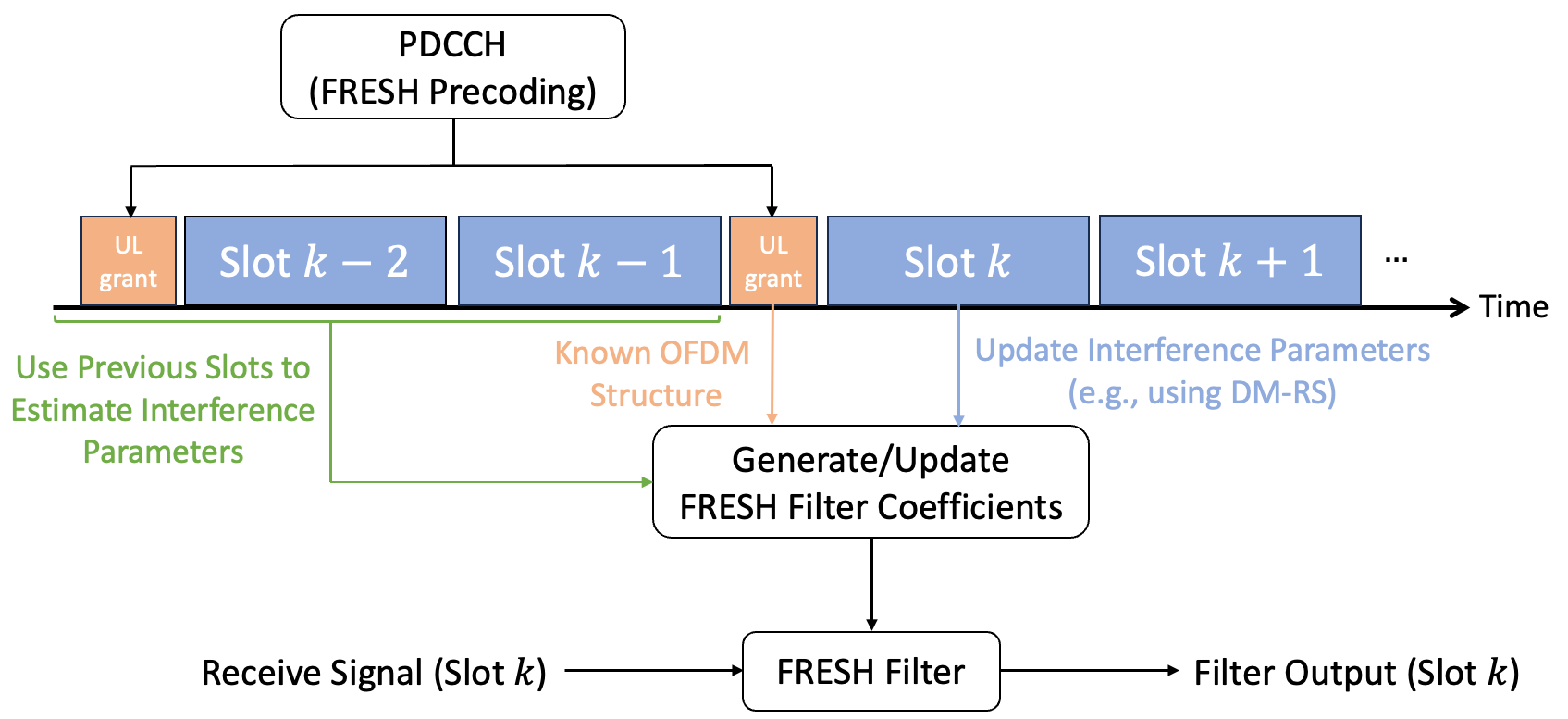}
\caption{Practical FRESH filter coefficient generation and update procedure for uplink OFDM reception using known scheduling information and interference-parameter estimates from received slots. }
\label{fig:5G_UL_FRESH}
\end{center}
\end{figure}

\subsection{Practical Implementation Considerations}
The analytical formulation developed above also enables a practical implementation in which a substantial portion of the FRESH filter coefficient generation procedure can be performed in advance. For example, in a 5G uplink system, as illustrated in Fig.~\ref{fig:5G_UL_FRESH}, the base station determines the time-frequency resources assigned to a scheduled user through the uplink resource grant. In an empirical FRESH filter, estimating the required second-order statistics would require a sufficiently long sequence of known or training symbols, introducing additional training overhead and reducing the resources available for data transmission. Moreover, if the interference characteristics change across subsequent slots, these statistics and the corresponding filter coefficients may need to be re-estimated and recalculated. In contrast, since the assigned subcarrier structure and the spectral redundancy introduced by the precoding scheme are known at the base station, the second-order statistics associated with the desired OFDM signal, including its SCF and cross-SCF terms, can be constructed directly from the corresponding precoding matrix. These quantities do not need to be re-estimated from the received waveform unless the precoding scheme changes. \par
The remaining quantities required for coefficient generation are primarily associated with the interference. For a pulse-shaped single-carrier interferer, for example, the analytical SCF is determined by parameters such as the symbol rate, carrier-frequency offset, and pulse-shaping associated parameters. These parameters can be estimated from received observations and reused across subsequent uplink transmissions as long as the interference characteristics remain sufficiently stationary. In particular, slowly varying parameters, such as the symbol rate and pulse-shaping roll-off factor, may be estimated over a relatively long observation interval, whereas parameters that are more susceptible to temporal variation, such as the interference frequency offset and received power, can be periodically updated. Known uplink reference-signal locations, such as DM-RS, can further facilitate this process by providing observations for which the desired-signal structure is known. \par

\section{Simulation Results}
\label{sec:V}
To evaluate the performance of the FRESH filter using the filter coefficients derived in Section~\ref{sec:III} and Section~\ref{sec:IV}, we present numerical results in this section. 

\subsection{Time-domain versus frequency-domain FRESH filtering}
\label{sec:VA}
We first compare the performance of the time-domain and frequency-domain implementations of the FRESH filter under a CP-free OFDM scenario, which is obtained by setting \(N_{\mathrm{cp}}=0\) for the time-domain FRESH filter. In this scenario, both implementations operate on the same untruncated interference signal and exploit the same frequency-domain redundancy. To distinguish the effect of the implementation domain from that of the convolution model, we additionally evaluate a frequency-domain linear-convolution implementation based on overlap-save processing. This serves as a reference for the block-wise circular implementation adopted in the remainder of the paper. This setup provides a common basis for selecting the time-domain filter length and frequency-domain FFT size such that the two implementations achieve approximately equivalent filtering performance. The selected parameters are then fixed in the next subsection for a consistent comparison between the pre-CP-removal and post-CP-removal FRESH filtering configurations. \par
\begin{figure}[t!]
\begin{center}
\includegraphics[scale = 0.45]{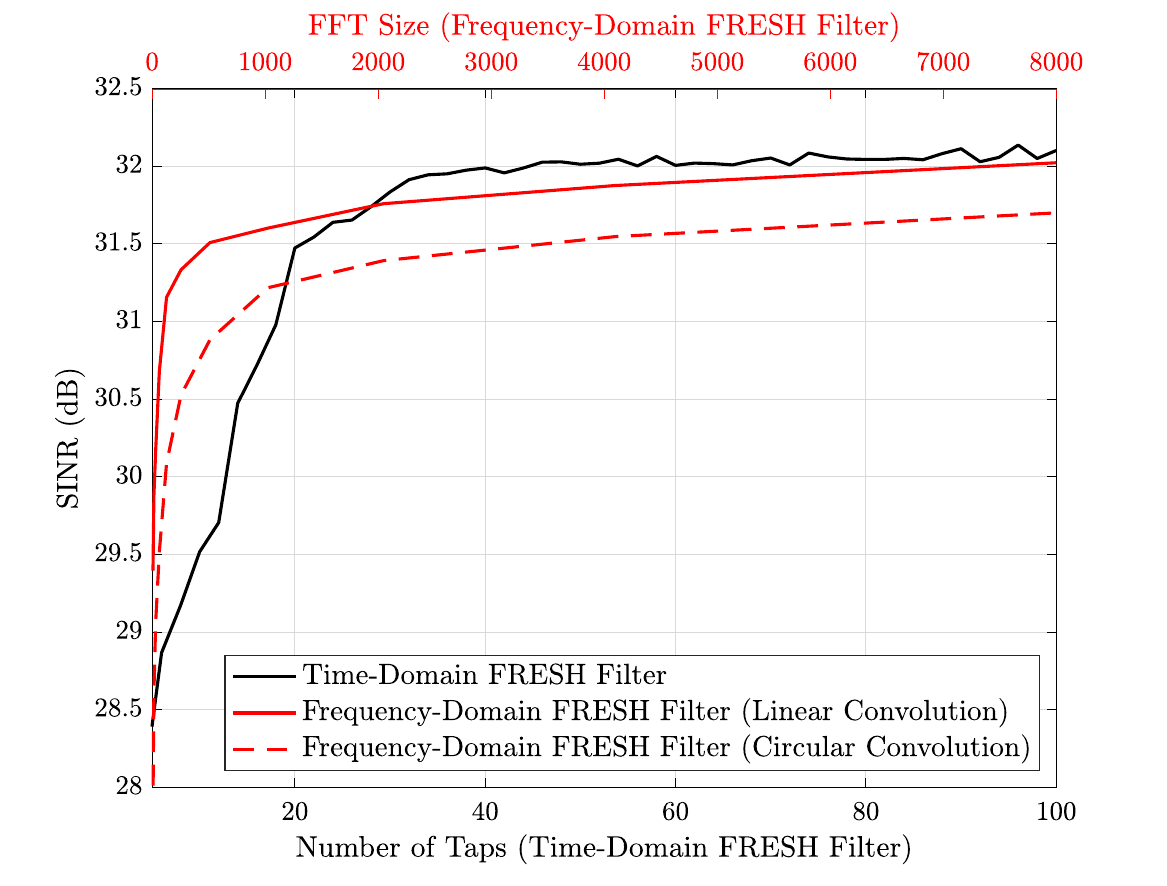}
\caption{Impact of the time-domain filter length and the frequency-domain FFT size on the post-filter SINR. SNR \(= 30\) dB, SIR \( = 0\) dB. }
\label{fig:time_vs_freq}
\end{center}
\end{figure}

As mentioned above, the desired signal is assumed to be a CP-free OFDM waveform with \(N\) subcarriers, where the frequency-domain OFDM symbol are repeated twice by setting the precoding matrix \(\mathbf{P} = \left[ I_{N/2} \  I_{N/2} \right]^T\), resulting in a frequency redundancy factor of \(N_{\mathrm{rep}} = 2\). The signal-to-interference ratio (SIR) is set to \(0\) dB. For this structure, the set of cyclic frequencies of the OFDM signal is \(\mathcal{A}_d = \left\{0, \frac{1}{2} \right\}\). Next, the interference signal is 16-QAM modulated and pulse shaped by a root-raised-cosine (RRC) filter with an upsampling factor \(P = 8\), a span of four symbols, and a roll-off factor of 0.5. The corresponding set of cyclic frequencies of the interference signal is \(\mathcal{A}_i = \left\{ -\frac{1}{2}, -\frac{1}{2} + \frac{1}{P}, \dots, \frac{1}{2} - \frac{1}{P}\right\}\). Therefore, the complete set of cyclic frequencies used in the FRESH filter corresponds to \(\mathcal{A}_i\). Fig.~\ref{fig:time_vs_freq} illustrates the impact of the number of taps in the time-domain FRESH filter and the number of FFT bins in the frequency-domain implementation, where the post-filter SINR is calculated as
\begin{equation}
\text{post-filter SINR} = \frac{\mathbb{E}\left[|d(t)|^2 \right]}{\mathbb{E}\left[\left|\hat{d}(t) - d(t) \right|^2\right]},
\end{equation}
representing the ratio of the desired signal power to the power of the residual estimation error. \par
As shown in Fig.~\ref{fig:time_vs_freq}, the time-domain FRESH filter and the frequency-domain implementation based on overlap-save exhibit similar performance and approach nearly the same post-filter SINR, as both correspond to linear convolution. In contrast, the circular-convolution implementation yields a lower SINR because each OFDM symbol is treated as a periodic block, introducing wrap-around at the block boundaries and resulting in time-domain aliasing. As the FFT size increases, the relative impact of the boundary effect decreases, and the circular-convolution performance converges to within approximately \(0.3\) dB of the linear-convolution implementation. The post-filter SINR reaches approximately \(90\%\) of its maximum observed value at around \(25\) taps for the time-domain implementation and \(1024\) FFT bins for both frequency-domain implementations, where the maximum value refers to the highest SINR obtained within the considered parameter range. Despite the slight performance loss, circular convolution is adopted in the remainder of the simulation results because it operates directly on the OFDM frequency-domain symbols available from the conventional receiver FFT and preserves the low-complexity structure discussed in Section~\ref{sec:IVC}. Fig.~\ref{fig:time_vs_freq} also indicates that equivalent performance can be achieved by appropriately selecting the filter length and FFT size, and the corresponding parameters are used in the following subsection for a consistent comparison. 
\par

\subsection{FRESH filtering for CP-OFDM and CP-removed OFDM} 
\label{sec:VB}
The time-domain FRESH filtering approach has the same implementation complexity regardless of whether the OFDM signal contains a CP. In contrast, the frequency-domain FRESH filtering approach can achieve significantly lower implementation complexity when applied to CP-free OFDM signals, since the SCF involving the desired signal for obtaining the optimal filter coefficients can be computed directly from the precoding matrix \(\mathbf{P}\). \par
\begin{figure}[t!]
\begin{center}
\includegraphics[scale = 0.45]{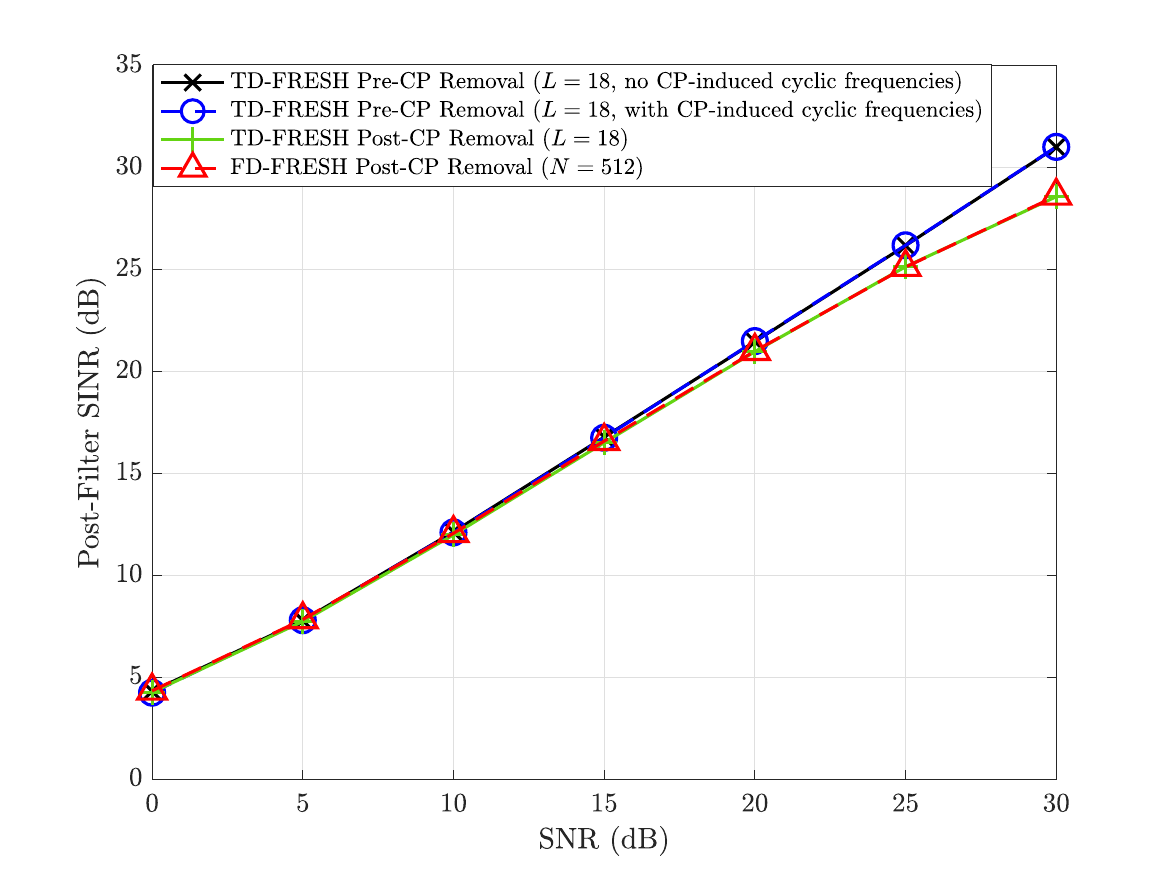}
\caption{SNR (per sample) vs. post-filter SINR comparison for the OFDM signals with and without CP using the FRESH filter. \(\text{SIR} = 0\) dB. }
\label{fig:cp_vs_nocp}
\end{center}
\end{figure}

In Fig.~\ref{fig:cp_vs_nocp}, we compare the interference cancellation performance for OFDM signals under four scenarios: 

\begin{enumerate}
    \item[(1)] Time-domain FRESH filtering of CP-OFDM signal without CP-induced cyclic frequencies (black curve).
    \item[(2)] Time-domain FRESH filtering of CP-OFDM signal with CP-induced cyclic frequencies (blue curve).
    \item[(3)] Time-domain FRESH filtering of CP-removed OFDM signal (green curve).
    \item[(4)] Frequency-domain FRESH filtering of CP-removed OFDM signal (red curve).
\end{enumerate}

The comparison between (1) and (2) evaluates the impact of the additional cyclic frequencies introduced by the CP on FRESH filtering performance. The comparison between (1)\(/\)(2) and (3) evaluates the effect of CP removal, which truncates the interference signal. Finally, the comparison between (3) and (4) evaluates the performance difference between time- and frequency-domain FRESH filtering for CP-removed OFDM signals. Based on the results in Section~\ref{sec:VA}, we select \(L = 18\) taps and \(N = 512\) bins for the time- and frequency-domain implementations, respectively. Although larger filter lengths and FFT sizes provide marginal performance improvements, these settings retain approximately \(85\%\) of the corresponding maximum observed post-filter SINR while providing a favorable tradeoff between filtering performance and computational complexity. All other parameters remain the same as in the previous subsection. The CP length is approximately \(7\%\) of the total number of subcarriers. Unlike CP-free OFDM, the presence of CP introduces additional cyclic frequencies. Specifically, the CP generates cyclic frequencies at \(\alpha = \frac{k}{N_s}\), where \(k = 0,1,\dots,N_s - 1\). Consequently, the set of cyclic frequencies for the desired signal includes not only the dominant cyclic frequencies \(\left\{0, \frac{1}{2}\right\}\), but also the CP-induced cyclic frequencies \(\left\{\frac{k}{N_s}: k = 0, 1,\dots, N_s-1 \right\}\). \par
Fig.~\ref{fig:cp_vs_nocp} shows that the interference cancellation performance of time-domain FRESH filtering for CP-OFDM is essentially identical regardless of whether the CP-induced cyclic frequencies are included. This indicates that the cyclostationarity introduced by the CP is negligible compared to that generated by the repetition of the frequency-domain symbols. This observation is further supported by the CAF shown in Fig.~\ref{fig:CAF_redun2}, where the tones around lag \(\tau = \pm N\) are significantly weaker than the dominant tones centered at \(\tau = 0\). On the other hand, the comparison between time-domain FRESH filtering of CP-OFDM and CP-removed OFDM signals in Fig.~\ref{fig:cp_vs_nocp} shows that interference truncation due to CP removal has a limited impact on the interference cancellation performance. The two approaches achieve similar performance at low SNR, while the CP-removed OFDM signal exhibits a performance degradation of approximately \(2\) dB at an SNR of \(30\) dB compared with the CP-OFDM signal. This indicates that the sub-optimal filter coefficients obtained under the assumption of uniformly distributed interference symbol timing within an OFDM block remain effective. Furthermore, the time- and frequency-domain implementations for CP-removed OFDM achieve comparable results. Since all four cases in Fig.~\ref{fig:cp_vs_nocp} provide similar interference cancellation performance, and both approaches require synchronization, the frequency-domain FRESH filtering after CP removal is adopted in the remaining simulations due to its lower implementation complexity. Specifically, the FFT is already available in a conventional OFDM receiver, allowing the FRESH filter to be implemented as a frequency-domain multiplication, whereas the time-domain implementation requires additional convolution operations. \par

\subsection{Frequency-domain FRESH filtering}
In this subsection, we evaluate the performance of the frequency-domain FRESH filter when the desired signal follows a 5G-like OFDM waveform. The desired signal contains \(N = 1024\) subcarriers with a frequency redundancy of \(N_{\mathrm{rep}} = 2\). A DC null subcarrier is placed at the center, with 306 repeated data subcarriers on each side, and 206 and 205 null subcarriers at the left and right edges, respectively. For this structure, the set of cyclic frequencies of the OFDM signal is \(\mathcal{A}_d = \{0, \pm 307/N\}\). The interference signal has the same structure described in Section~\ref{sec:VA}. 
To compare with the conventional interference mitigation approaches, we also consider applying the Wiener filter to the received signal. This corresponds to a special case of the FRESH filter in which only the cyclic frequency \(\alpha = 0\) is considered. \par

\begin{figure}[t!]
\begin{center}
\includegraphics[scale = 0.45]{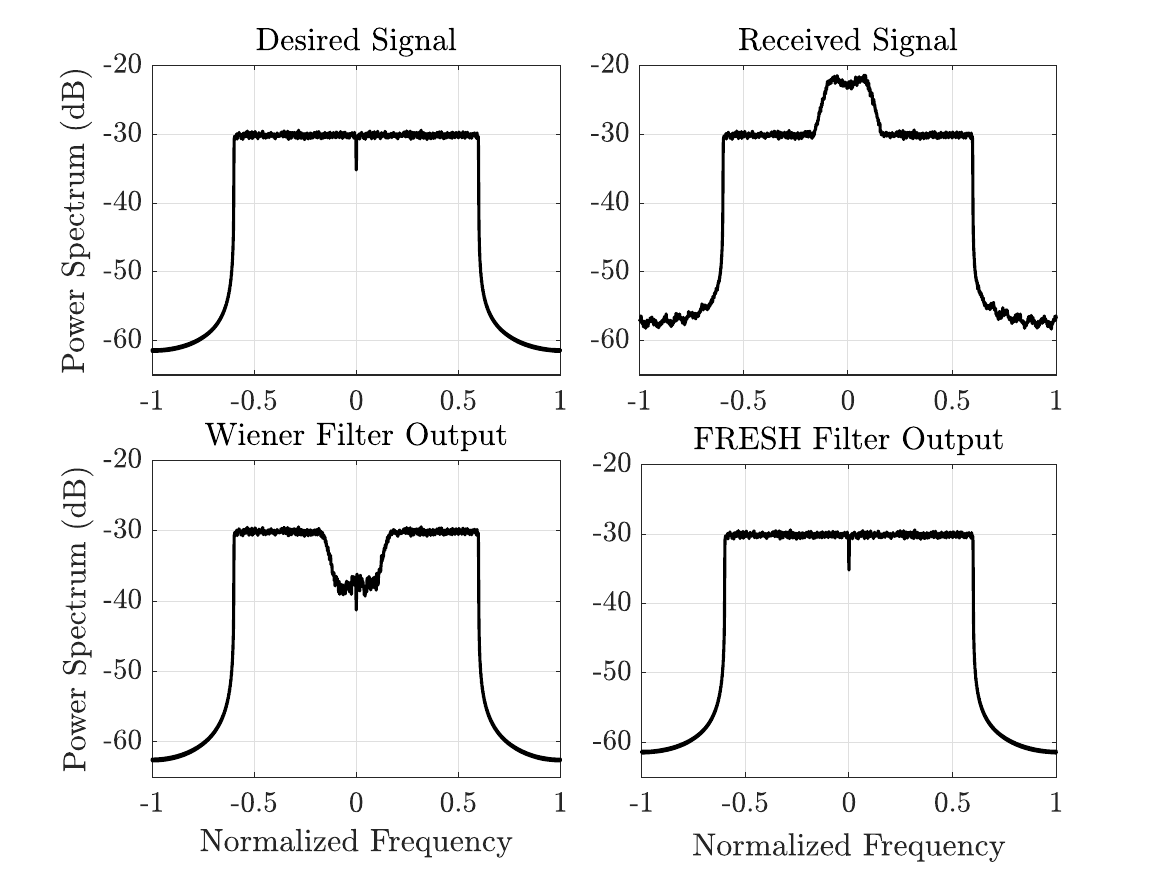}
\caption{Power spectrum of the received signal and the desired signal estimates from the Wiener filter and the FRESH filter. The interference is a single-carrier signal centered within the desired signal bandwidth. \(\mathrm{SNR} = 30\) dB, \(\mathrm{SIR} = 0\) dB. }
\label{fig:psp}
\end{center}
\end{figure}

\begin{figure}[t!]
\begin{center}
\includegraphics[scale = 0.55]{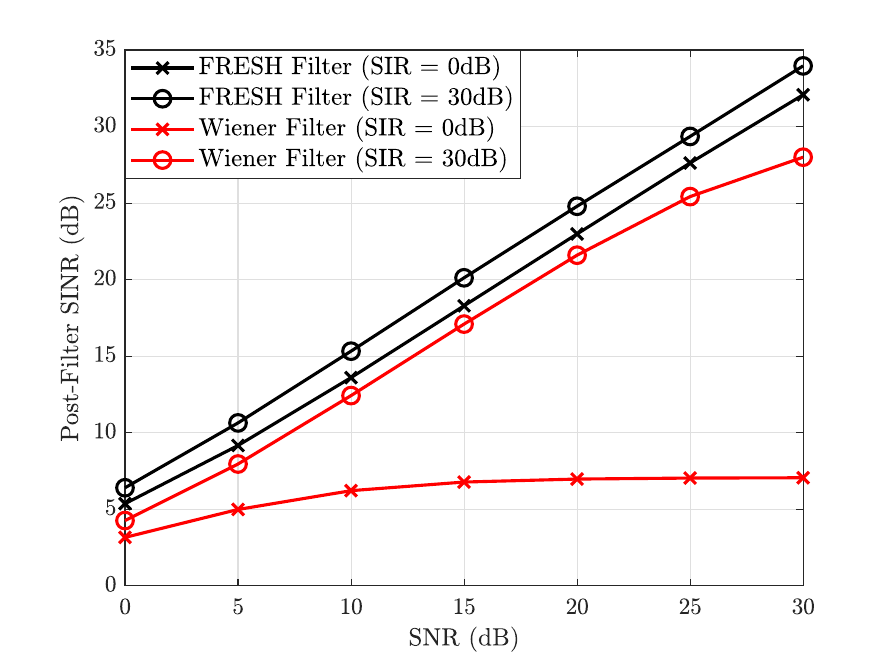}
\caption{SNR (per sample) vs. post-filter SINR comparison for the FRESH filter and the Wiener filter applied to an OFDM signal with a frequency-domain redundancy factor of two. }
\label{fig:snr_vs_sinr}
\end{center}
\end{figure}

To illustrate the effect of filtering on the desired signal and interference, Fig.~\ref{fig:psp} compares the power spectrum of the desired signal estimates obtained using different filters. The Wiener filter strongly attenuates the interference-contaminated spectral region, but this also suppresses the overlapping desired-signal components, resulting in a spectral scooping. Because the interference power is concentrated within a relatively narrow bandwidth, the local SINR in the overlapping region is low, causing the Wiener filter to assign a small gain to these frequencies. In contrast, the FRESH filter constructs a weighted combination of the frequency-shifted spectral components and exploits the inherent spectral redundancy of the desired signal, enabling effective interference suppression while largely preserving the desired-signal spectrum. This behavior is further reflected in Fig.~\ref{fig:snr_vs_sinr}, which compares the post-filter SINR over a range of input SNR and SIR values. When the interference is strong (\(\mathrm{SIR}=0\) dB), the spectral scooping introduced by the Wiener filter causes the post-filter SINR to saturate as the input SNR increases. In contrast, the FRESH filter preserves the desired signal while suppressing the interference, allowing the post-filter SINR to continue increasing with SNR. When the interference is weaker (\(\mathrm{SIR}=30\) dB), the distortion caused by the Wiener filter becomes less pronounced, and both methods exhibit increasing post-filter SINR, although the FRESH filter consistently achieves better performance.  \par

\begin{figure}[t!]
\begin{center}
\includegraphics[scale = 0.45]{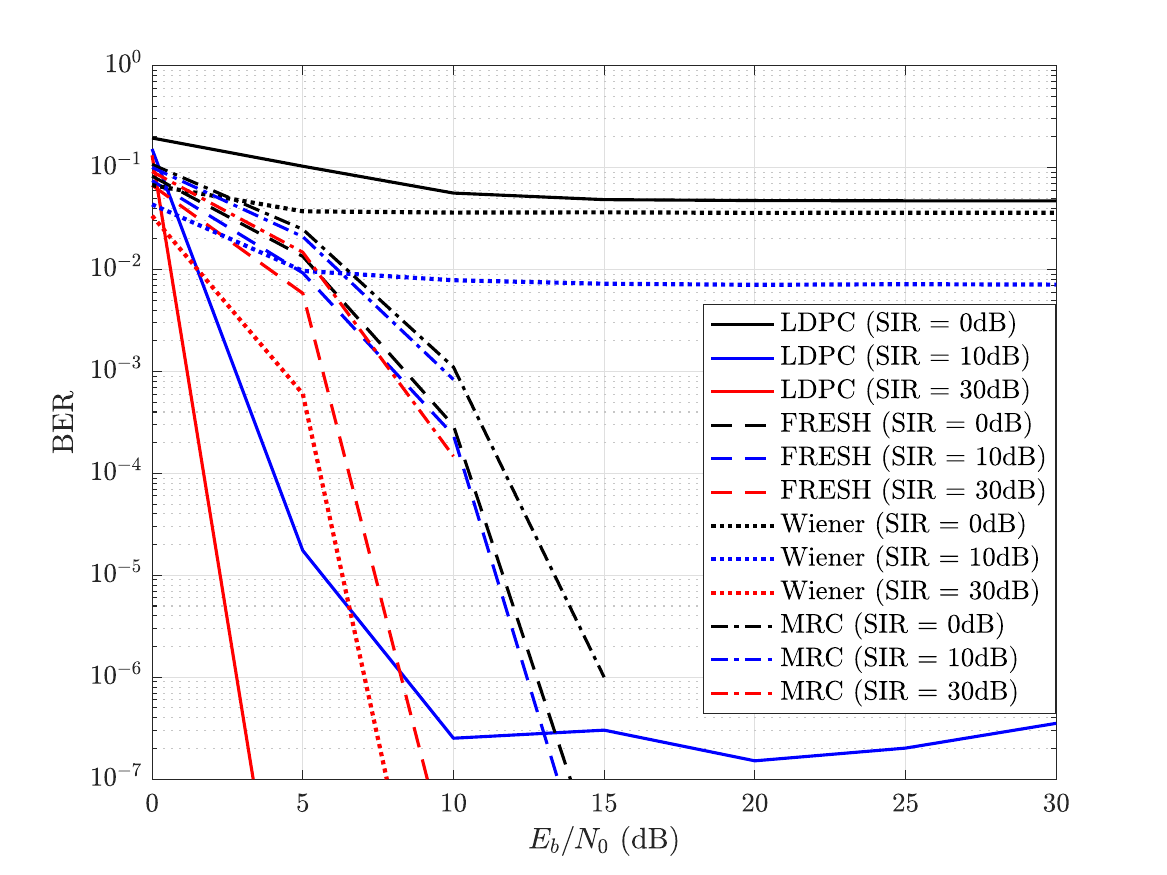}
\caption{SNR per information bit vs. BER comparison for LDPC coding (\(1/2\)-rate code, \(16\)-QAM), FRESH filter (repetition-\(2\), \(16\)-QAM), Wiener filter (no repetition, QPSK) and MRC (repetition-\(2\), \(16\)-QAM), all configured to achieve the same spectral efficiency of \(2\)b/s/Hz. }
\label{fig:EbN0_vs_BER}
\end{center}
\end{figure}

Building upon the previous comparison in terms of the estimation accuracy of the desired signal, we now evaluate the performance from a bit-error-rate (BER) perspective by incorporating channel coding. Specifically, we employ a DVB-S2 LDPC code with rate \(1/2\), consisting of \(32400\) information bits and a \(64800\)-bit codeword. The encoded bits are directly mapped to the symbols without additional interleaving. For comparison, we also consider the frequency-domain maximum ratio combining (MRC) using the same repetition factor of two as the FRESH filter and LDPC precoding scheme. In the MRC implementation, after applying the FFT to the received signal, the two repeated observations, \(X_{k}\) and \(X_{k+N/2}\), are combined according to their interference-plus-noise powers as
\begin{equation}
\hat{X}_k = \frac{X_{k}/\sigma_{k}^2+X_{k+N/2}/\sigma_{k+N/2}^2}{1/\sigma_k^2+1/\sigma_{k+N/2}^2}, \quad k = 1, 2, \dots, N/2,
\end{equation}
where \(\sigma_k^2\) and \(\sigma_{k+N/2}^2\) denote the interference-plus-noise powers at subcarriers \(k\) and \(k+N/2\), respectively. To ensure a fair comparison at the same spectral efficiency, the LDPC, FRESH filter, and MRC schemes use \(16\)-QAM, while the Wiener filter, which does not require repetition redundancy, uses QPSK. The BER results in Fig.~\ref{fig:EbN0_vs_BER} show that under strong interference, LDPC coding alone performs poorly, as it is primarily designed for noise-limited channels rather than interference-dominated scenarios. When \(\mathrm{SIR} = 0\)~dB, the Wiener filter slightly outperforms standalone LDPC coding because it partially suppresses the interference. In contrast, MRC achieves a lower BER than both methods, while the FRESH filter further improves the BER performance, particularly in the high-SNR regime. When the interference is weak (\(\mathrm{SIR = 30}\) dB), LDPC coding achieves an extremely low BER for \(E_b/N_0 \ge 5\) dB, as the channel now becomes predominantly noise-limited. In this regime, reducing the interference power yields only marginal BER improvement for the FRESH filter and the MRC approaches, while the Wiener filter—operated using QPSK modulation without any coding redundancy—experiences a substantial BER reduction. These results indicate that FRESH filtering is particularly beneficial in interference-dominated environments, whereas LDPC coding excels in noise-limited conditions. A more practical strategy is therefore to combine the two approaches: when interference is weak, directly applying LDPC coding is sufficient; when interference is strong, one may first employ the FRESH filter for interference suppression, followed by a higher code-rate error-correction coding scheme to maintain the overall throughput. \par

\begin{figure}[t!]
\begin{center}
\includegraphics[scale = 0.45]{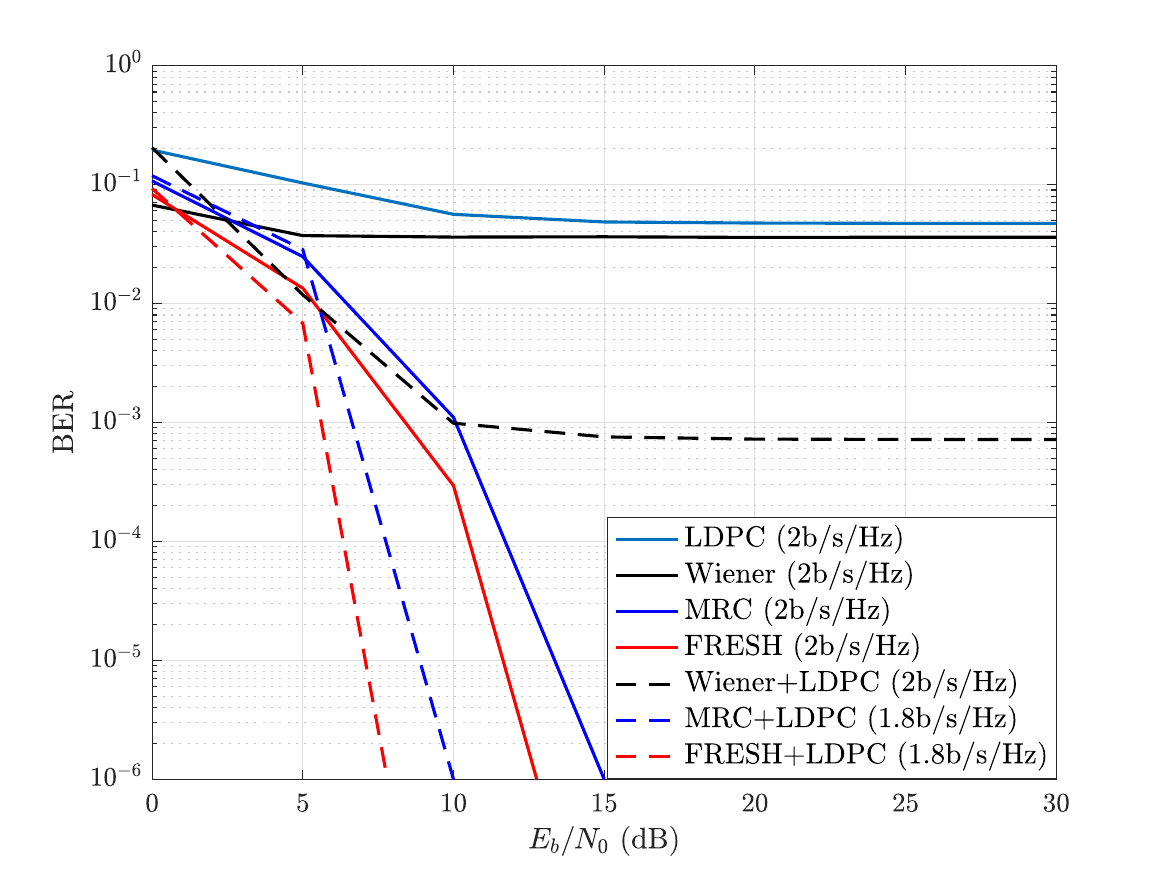}
\caption{SNR per information bit vs. BER comparison for LDPC coding (16-QAM), FRESH filter (16-QAM), Wiener filter (QPSK), and MRC (16-QAM), as well as the combined Wiener/MRC/FRESH+LDPC (16-QAM) schemes. }
\label{fig:EbN0_vs_BER_full}
\end{center}
\end{figure}
To validate this strategy, Fig.~\ref{fig:EbN0_vs_BER_full} compares the BER performance of LDPC-only decoding, standalone Wiener filtering, FRESH filtering, and MRC, together with their respective combinations with LDPC decoding, under an SIR of \(0\) dB. For the LDPC/FRESH/MRC-only cases, the effective rate is \(0.5\). In the combined schemes, the Wiener filter is followed by a rate-\(1/2\) LDPC code for a fair comparison, whereas the FRESH filter and MRC each operate at a rate of \(0.5\) and are followed by a rate-\(0.9\) LDPC code, so that the overall spectral efficiency remains approximately unchanged. Under strong interference, combining Wiener filtering with error-correction coding provides a noticeable BER reduction. However, the BER eventually exhibits an error floor around \(10^{-3}\), which is due to the limited interference-suppression capability of the Wiener filter. In contrast, the MRC+LDPC scheme achieves a substantial BER improvement for \(E_b/N_0 \geq 10\) dB. The FRESH+LDPC scheme provides further performance gains, with a pronounced BER reduction beginning at approximately \(E_b/N_0 \geq 5\) dB. 

\section{Conclusion}
\label{sec:VI}
This paper developed a FRESH filtering framework for OFDM systems operating in the presence of a co-channel interferer. By characterizing the second-order statistics of OFDM signals with and without CP, we derived compact closed-form matrix expressions for the optimal FRESH filter coefficients. The proposed formulation enables offline coefficient computation and efficient frequency-domain implementation, making it well suited to practical OFDM receivers. The analytical formulation also enables reuse of precomputed desired-signal statistics and lightweight coefficient updates when only a subset of interference parameters varies. Simulation results demonstrated that the FRESH filter significantly outperforms the conventional Wiener filter and standalone LDPC coding in interference-dominated scenarios. By exploiting the spectral redundancy and cyclostationary structure, the FRESH filter effectively suppresses strong co-channel interference while introducing minimal signal distortion. Moreover, the results also highlight that interference suppression and channel coding play complementary roles. When the interference is weak, standalone LDPC coding is sufficient as the system becomes noise-limited. In contrast, under strong interference, the combination of FRESH filtering and higher-rate LDPC coding achieves the lowest BER among the considered schemes while preserving the overall throughput.

\section*{Acknowledgment}

The authors declare that no artificial intelligence (AI) or generative AI tools were used in conducting the research, performing the analysis, or preparing this manuscript.

\bibliographystyle{IEEEtran}
\bibliography{references}

\vspace{11pt}

\end{document}